\documentclass{cernrep}
\usepackage{color}
\usepackage{amssymb} %added by GP
\usepackage{amsmath}  %added by GP
\usepackage{slashed}  %added by GP
\usepackage[normalem]{ulem} 
\usepackage {ulem}
\begin{document}
\title{Photoproduction with a mini-jet model and Cosmic Ray showers}
\author{  F. Cornet  $^1$,  C. Garcia-Canal  $^2$, A. Grau  $^1$, G. Pancheri $^3$ \thanks{ Research affiliate with MIT CTP, Cambridge, MA, USA},
S. J. Sciutto  $^2$ } %On leave from another institue somewhere.}}
\institute{
%\institute{
$^1$ Departamento de F\'{\i}sica Te\'orica y del Cosmos,  
%and Centro Andaluz de F\'{\i}sica de Part\'{\i}culas, 
Universidad de Granada, E-18071 Granada, Spain \\
$^2$ Departamento de F\'{\i}sica, Universidad Nacional de La
Plata, IFLP, CONICET, C. C. 67, 1900 La Plata, Argentina\\
$^3$ INFN Frascati, National Laboratories,  Frascati  00044, Italy
%\institute
}
%\institute{
%\author{A. Grau}% \thanks
                % {Also research affiliate with MIT CTP, Cambridge, MA, USA}}
%                 On leave from another institue somewhere.}}
%\institute{Granada, Spain}
\begin{abstract}
We present post-LHC updates of estimates of the  total photo-production cross section in a mini-jet model with infrared soft gluon resummation, and  apply the model to study Cosmic Ray shower development, comparing the results with those obtained from other existing models.
%Each paper should be preceded by a short abstract of not more
%than 150~words, which should be written as a single paragraph
%and should not contain References and notes.
\end{abstract}


\maketitle

\section{The total photo production cross section}\label{sec:model}

We address again the question of how different models for photo production affect {the description of the development of photon-initiated} cosmic ray showers and, consequently,  how much the estimated photon composition of the showers depends from the models used in the simulation. In this way, we update a previous publication \cite{Cornet:2015qda}. The simulation of the shower development is performed using the AIRES MC \cite{AIRES}, linked to the hadronic models QGSJET-II-04 \cite{QGSJET-II-04}, QGSJET in the following, and EPOS-LHC  3.40 \cite{EPOS-LHC}, EPOS in the following,   that have recently %\textcolor{red}{been} 
been updated to take into account LHC results. The study includes two different photo-production models:
\begin {itemize}
\item the Block and Halzen (BH) model for total cross sections at very high energies ~\cite{Block:2004ek}	
 presently in AIRES as default model,
\item the extension to photoproduction ~\cite{Godbole:2008ex} of the so-called Bloch-Nordsieck (BN) model for total hadronic cross sections \cite{
Grau:1999em,Godbole:2004kx}, recently  implemented in AIRES.
\end{itemize}

\subsection{The Bloch-Nordsieck (BN)  model for
% {\textcolor{red}{purely}} 
hadronic processes} \label{ssec:BN}
\begin{figure}[ht]
%\begin{center}
\vspace{-4 cm}
\includegraphics[width=7cm]{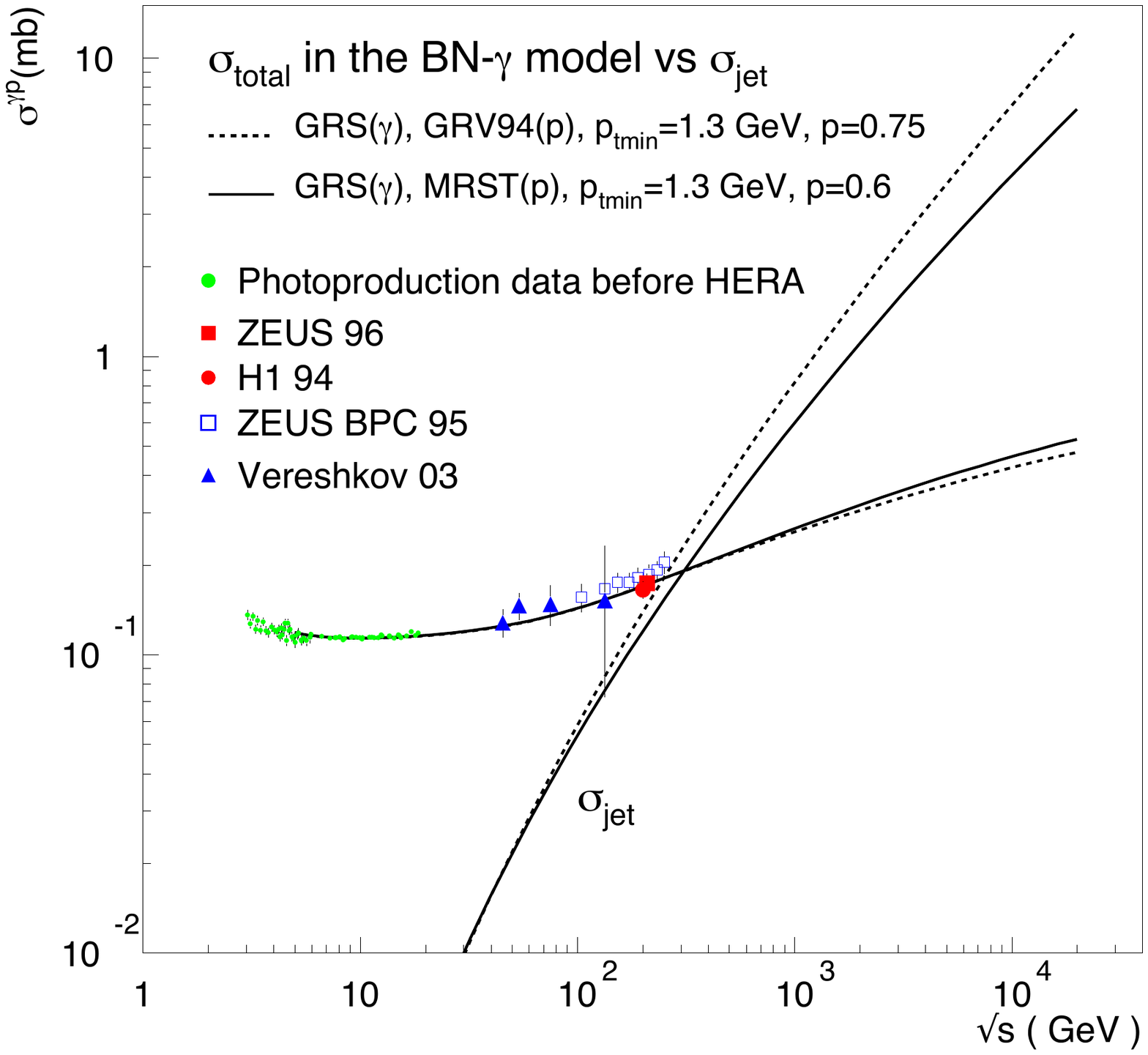}
\vspace{+1.5cm}
\vspace{-2.0cm}
\includegraphics[width=7cm]{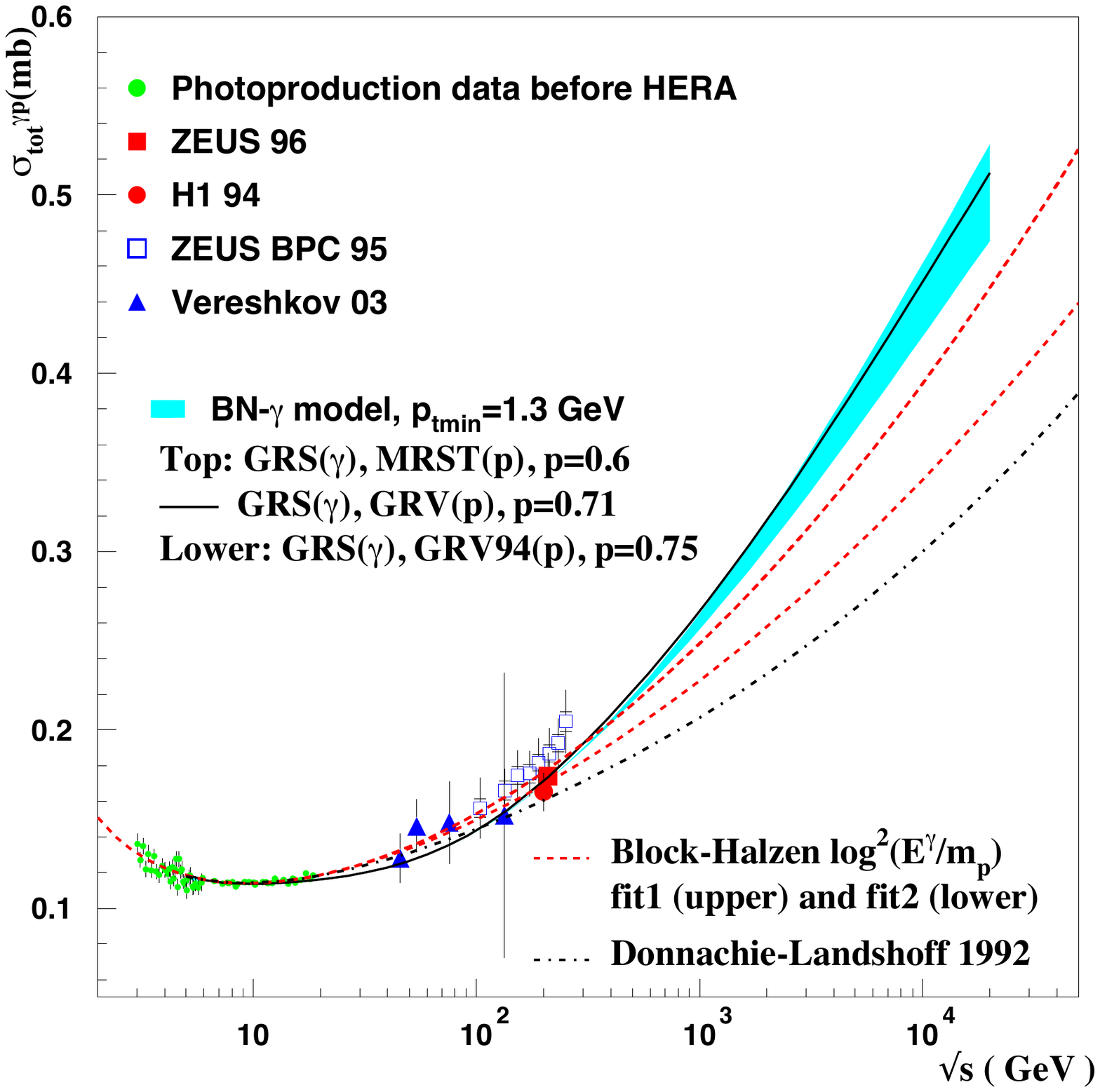}
%gamp_auger }
\vspace{-1.3cm}
\caption{Left panel: Mini-jet integrated cross sections for different PDFs  in $\gamma p$ scattering,  in comparison with $\gamma p$ total cross section data and the BN model described in the text. Right panel: Total $\gamma p$ cross section  calculated  for different PDFs with    the  BN model, called here BN-$\gamma$,  and compared   with the model of
 Ref.~\cite{Block:2004ek}, with lower red dashed curve  being  the default in AIRES,  and the Donnachie and Landshoff description of Ref.~\cite{Donnachie:1992ny}.}
\label{fig:sigjet_sigtot_gamp}
%\end{center}
\end{figure}

This model for the total hadronic cross section is based on a perturbative QCD (pQCD) calculation of mini-jets as being at the origin of the observed rise \cite{Durand:1988ax} of the total cross section with energy.  For  a fixed minimum transverse momentum of the scattering partons, called $p_{tmin}\gtrsim 1$ GeV, the low-x behavior of the PDFs leads to  a very strong increase  of the minijet integrated cross sections as  the c.m. energy increases,  shown in the left  panel of Figure~\ref{fig:sigjet_sigtot_gamp}
for  different LO Parton Density Functions (PDFs): Gl\"uck, Reya and Schienbein (GRS)  \cite{GRS} for the photon, Gl\"uck, Reya and Vogt (GRV)  \cite{GRV}, and Martin, Roberts, Stirling and Thorne (MRST) \cite {MRST} for the proton. In the BN model, to be described shortly,  infrared gluon resummation  tames the fast rise of mini-jet cross sections  through  soft gluon emissions
% in the infrared region 
and can lead to  saturation through    a phenomenological  ansatz for resummation of $k_t\simeq 0$ gluons, {which we outline below}. Thus the sudden     rise around $\sqrt{s} \gtrsim  10-20$ GeV morphs into  the   experimentally observed   gentle asymptotic behaviour,  which satisfies  the Froissart-Martin  bound, i.e. $\sigma_{total}\lesssim [\log s]^2$. 
To this effect, mini-jet collisions  are embedded into the eikonal formulation for the total cross section, {which, for $pp$ scattering, reads as }
\begin{equation}
\sigma_{total}=2\int d^2 {\bf b} [1-e^{-[  {\bar n}_{soft}(b,s)+{\bar n}_{hard}(b,s)]/2}] \label{eq:sigtot}
\end{equation}
{Here ${\bar n}_{hard}(b,s)$ is to be calculated through the LO, DGLAP evolved QCD jet cross section, while  our choice for  } ${\bar n}_{soft}$ 
%\sout{where fully non-perturbative effects take place,}  
is to  parametrize it by normalizing $\sigma_{total}$ at low energy, i.e.  before pQCD mini-jet  production takes up a major role.   Phenomenology suggests that this quantity  is either a constant  or decreases with energy. On the other hand, the main point of the model used here  is that ${\bar n}_{hard}(b,s)$ should be fully estimated by means of a pQCD calculation, with  saturation effects due to  an All Order Resummation procedure, which includes  the infrared region.

Once pQCD can be applied, a  complete description requires not only the calculation of hard parton-parton scattering but also   soft gluon effects accompanying the collision.  If, in Eq.~(\ref{eq:sigtot}), we  write  
\begin{equation}
{\bar n}_{hard}(b,s)=A_{hard}(b,s) \sigma_{jet}(s;p_{tmin},PDFs)\label{eq:nbs}
\end{equation}
then  $A_{hard}(b,s)$  will  include soft gluon resummation effects, and thus account for the cut-off in impact parameter space, required for satisfaction of the Froissart bound,  i.e. the   saturation effects.  

%\sout{At the same time, $ \sigma_{jet}(s;p_{tmin},PDFs)$ accounts for the rise: the two quantities however are interdependent, as  clearly seen in  Figure\ref{fig:sigjet_sigtot_gamp}, where the large discrepancy between calculation of  $\sigma_{jet}$  at asymptotic  energies due to different small-x behavior  of the PDFs   washes out in the full total cross section eikonalization.
%\subsubsection{A democratic resummation process for infrared gluons} \label{sss:resume}Saturation effects are obtained through formulation of the problem in the eikonal expansion, and use of an appropriate ansatz for the needed cut-off in impact parameter space. The physics behind our model for saturation relies on the acollinearity introduced by soft gluon emission always accompanying any hard or semi-hard parton-parton collision. }

 Our model for resummation of soft gluons is based on a semi-classical approach in which one does not count individual gluons, hence no branching or angular ordering is involved, and follows the Bloch and Nordsieck observation \cite{Bloch:1937pw} that  soft photons emission follows a Poisson distribution and only an infinite number of them can give a finite cross section. The procedure to apply this result to QCD
 {was first discussed in \cite{Corsetti:1996wg}} and  recently outlined in Ref ~\cite{Fagundes:2017xli}. Labelling as $A_{BN}$ the impact parameter distribution of partons to use in Eq.~(\ref{eq:nbs}),  the resulting expression is  the following: 
 %\cite{Corsetti:1996wg}:
\begin{eqnarray}
A_{BN}(b,s;p,p_{tmin})=\frac{
e^{-h(b,s;p,p_{tin}})
}
{
\int d^2{\bf b}e^
{-h(b,s;p,p_{tmin})}
}
 \equiv \\
\mathcal{N}(s,p_{tmin}) \int d^2{\bf  K_t} e^{-i{\bf K}_t\cdot {\bf b}}
 \frac {d^2P_{soft-resum}({\bf K_t},s;p_{tmin})}{d^2{\bf  K_t}}\\
 \label{eq:ABN}
\end{eqnarray} 
with
\begin{eqnarray}
h(b,s;p,p_{tmin})=\frac{8}{3\pi^2}\int _0^{q_{max}} d^2 {\bf k}_t [
    1-e^{i{\bf  k}_t\cdot {\bf  b} }]\alpha_s(k_t^2)\frac{\ln (2 q_{max}/k_t)}{k_t^2} \label{eq:hb}\\
q_{max}(s;p_{tmin},PDF)= \frac{\sqrt{s}} {2} 
\frac{ \sum_{i,j}\int \frac{dx_1}{ x_1} f_{i/a}(x_1)\int \frac{dx_2}{x_2}f_{j/b}(x_2)\sqrt{x_1x_2} \int_{z_{min}}^1 dz (1 - z)}
{\sum_{i,j}\int \frac{dx_1} {x_1}
f_{i/a}(x_1)\int \frac{dx_2} {x_2} f_{j/b}(x_2) \int_{z_{min}}^1 (dz)}\label{eq:qmax}
%q_{max}(s;p_{tmin},PDF)= 
%expression\  to\  be \ input\  by \  Agnese \label{eq:qmax}
\end{eqnarray} 
%with $z_{min}=4p_{tmin}^2/(sx_1x_2)$
{where the integration in Eq.~(\ref{eq:hb})  makes  }use of  an ansatz of maximal singularity for the behavior of the coupling of infrared gluons to the source current, namely 
$\alpha_s(k_t\rightarrow 0)\propto (k_t^2/\Lambda^2)^{-p}$ with $p<1/2<1$ \cite{Corsetti:1996wg}, and $q_{max}$ is calculated from the kinematics of single gluon emission \cite{Greco}.

%\subsection{Extending the BN model to photo production}
 In Reference~\cite{Godbole:2008ex} the model described above {was 
%in Subsect. \ref{sec:BNprotons} 
 applied to photon processes, using photon PDFs, and a suitable parametrization  for the probability $P_{had}$ that a photon behaves like a hadron, following the  simple, but  effective model  proposed in Ref.~\cite{Fletcher:1992mw}. In \cite{Godbole:2008ex} the total photo production cross section was then  calculated as follows:
\begin{equation}
\sigma_{total}^{\gamma p}=2 P_{had}\int d^2 {\bf b} [1-e^{-[2/3\  {\bar n}^{pp}_{soft}(b,s)+{\bar n}^{\gamma p}_{hard}(b,s)]/2}]
\end{equation}
%This model was used in \cite{Godbole:2008ex} 
with 
\begin{equation}
{\bar n}_{hard}^{\gamma p}(b,s)=\frac{
A^{\gamma p}_{BN}(b,s;p,p_{tmin},PDF)\sigma^{\gamma p}_{jet}(s; p_{tmin}, PDF) \label{eq:gamphard}
}{P_{had}}.
\end{equation}
%\sout{with Eq.~(\ref{eq:hard})  explicitly showing the dependence of the average number of hard collisions upon the model parameters. }
 The probability $P_{had}$ can be extracted from Vector Meson Dominance models, and adjusting it  to the normalization of data at low energy, we propose $P_{had}=1/240$. 
 
After determining the best set of parameters $\{p,p_{tmin}\}$ compatible with existing $\gamma p $ data,  we show in the right hand panel of Figure~\ref{fig:sigjet_sigtot_gamp}
%{\bf  I propose now to keep this figured}  
our present calculation \cite{Cornet:2015qda} for  the photo production total cross section, updated from \cite{Godbole:2008ex}  with  more recent  sets of proton  PDFs, 
%as in  from \cite{Cornet:2015qda},
 and compare it with the results of  the models from Refs.~\cite{Block:2004ek} and  \cite{Donnachie:1992ny}.
%{\bf  please check   about Block Ref.}. \begin{figure}[ht]\begin{center}\vspace{-2cm}\includegraphics[width=12cm]{gamp_auger}\vspace{-2cm}\caption{Total $\gamma p$ cross section  for different PDFs from  \cite{Cornet:2015qda} in  the  BN model described in the text in comparison with the model ofRef.~\cite{Block:2004ek}. {\bf Agnese will try to  
  %already been published. Perhaps we couldadd one Donnachie and Landshoff fit, just to make it different, for instance the 1992 version or 1996.} }\label{fig:gamptot}\end{center}\end{figure}

\section{Shower development with post LHC AIRES simulations}\label{sec:simulations}

%\section{Simulation Results}

\begin{figure}[htb]
	\begin{center}
		\begin{tabular}{cc}
			\includegraphics[width=7cm]{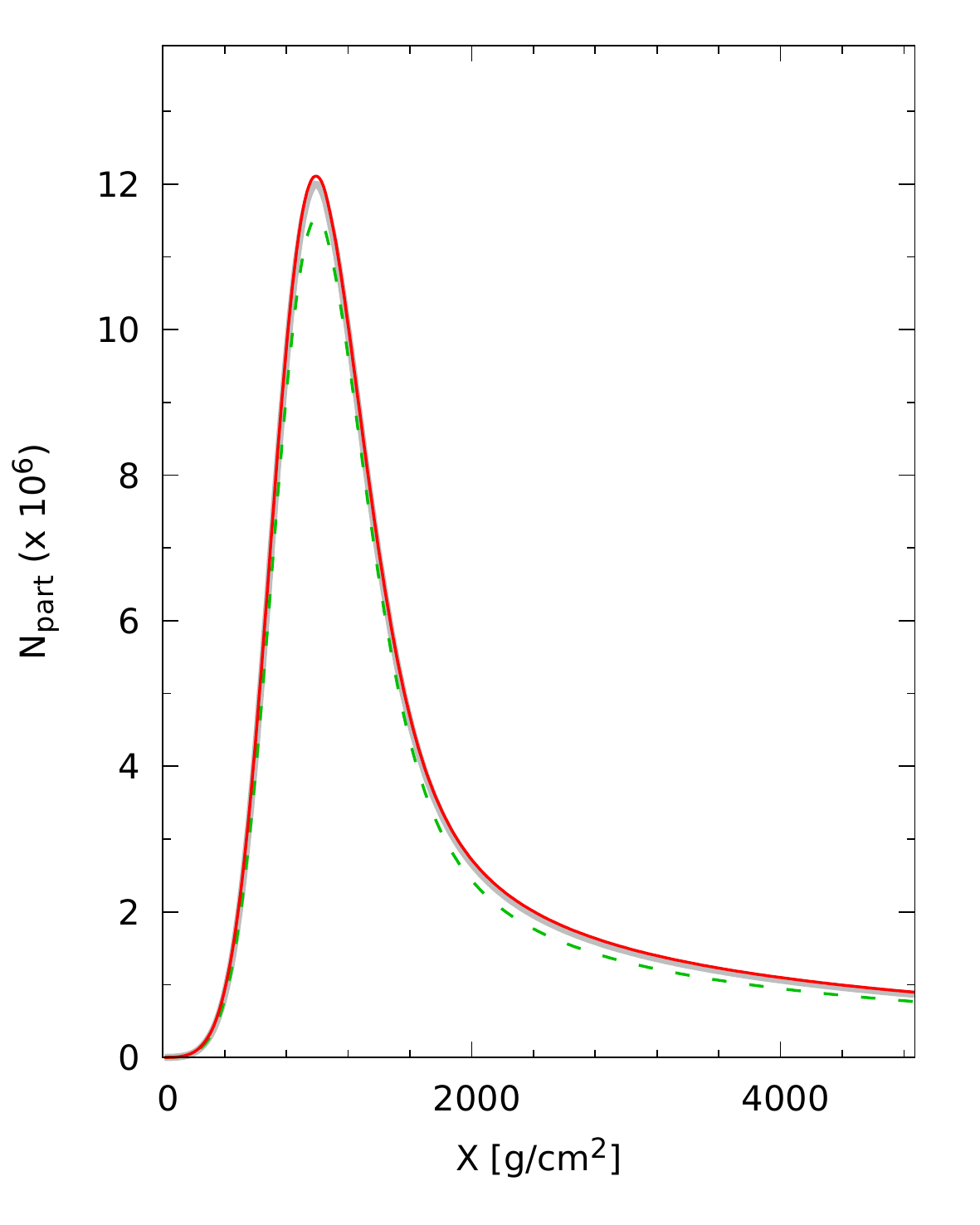} &
			\includegraphics[width=7cm]{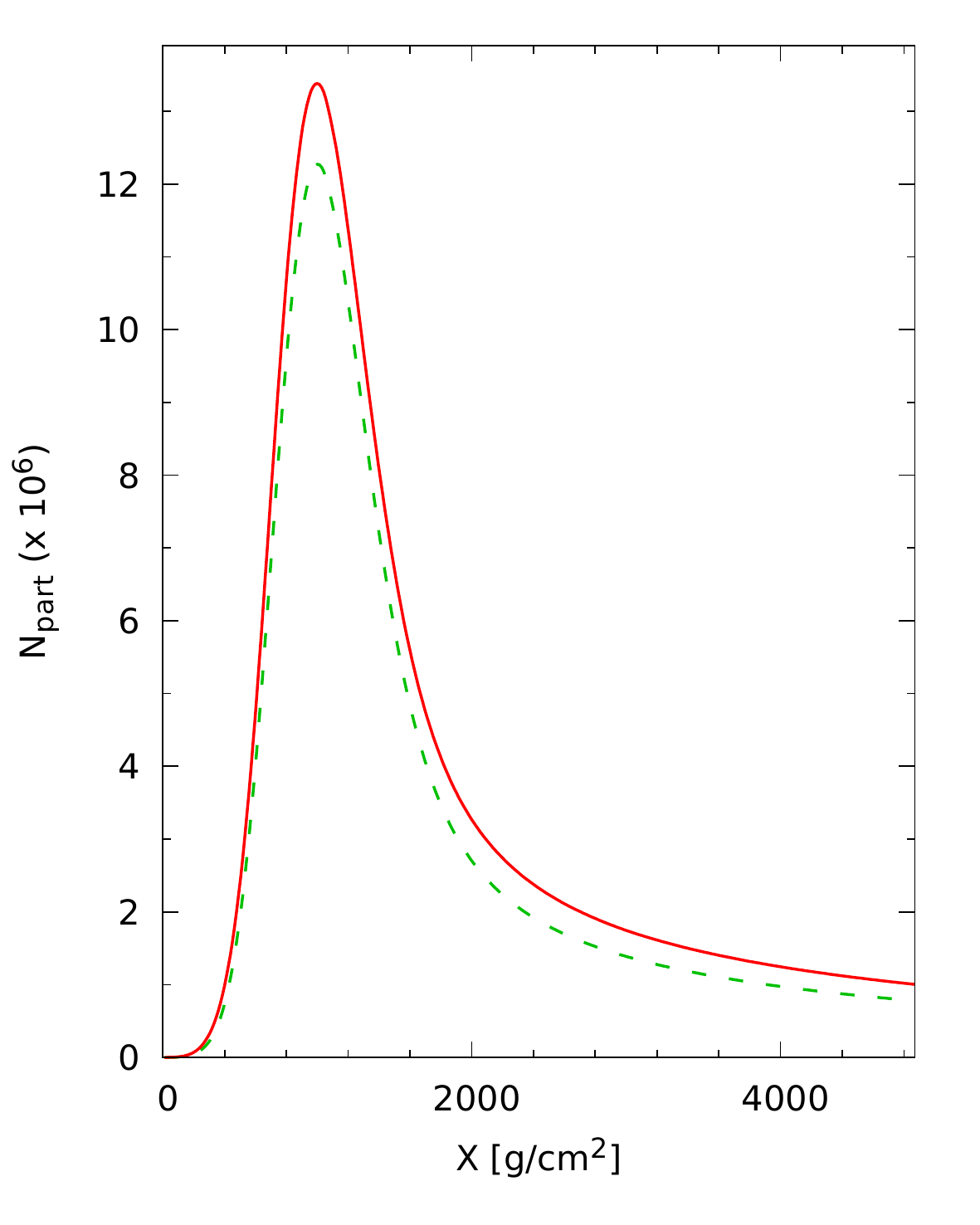}
		\end{tabular}
		\caption{Longitudinal development of muons for $10^{19}$ eV photon
			showers inclined 80 degrees.
			The solid (dashed) lines correspond to simulations with the
			BN-$\gamma$ (BH) model for photonuclear cross section, and processing high energy hadronic interactions with the QGSJET-II-04 (left) and EPOS (right) models. }
		%	{
		%\bf Lia: I have substituted the previous figures with the grey line with the figures without grey lines sent for the talk. If you do not agree, we can revert to to the first set of figures, in which case we keep the sentence : The grey line corresponds to similar simulations performed using the present model for photonuclear cross section and the (pre-LHC) QGSJET-II-03 model (see figure 5 of reference \cite{Cornet:2015qda}).}}
		\label{fig:longimuons}
	\end{center}
\end{figure}

\begin{figure}
	\begin{center}
       \includegraphics[width=7cm]{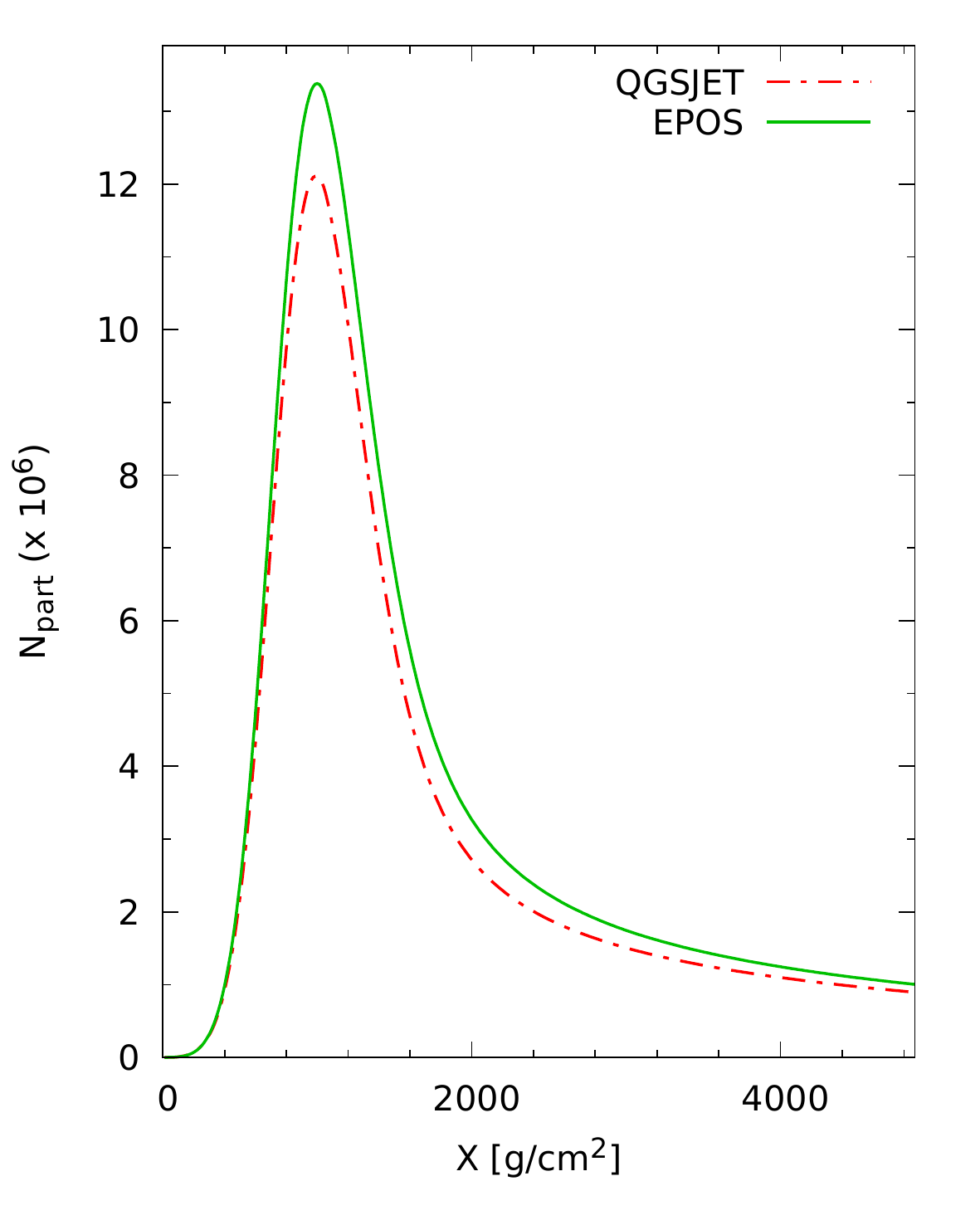}
         \caption{Longitudinal development of muons for $10^{19}$ eV photon
			showers inclined 80 degrees from QGSJET-II-04 or EPOS 3.40 hadronic packages in the case of the BN-$\gamma$ model for $\gamma p$.}
         \label{fig:compEPOSQGS}
     \end{center}
\end{figure}
\begin{figure}[tp]
	\begin{center}
		\begin{tabular}{cc}
			\includegraphics[width=7cm]{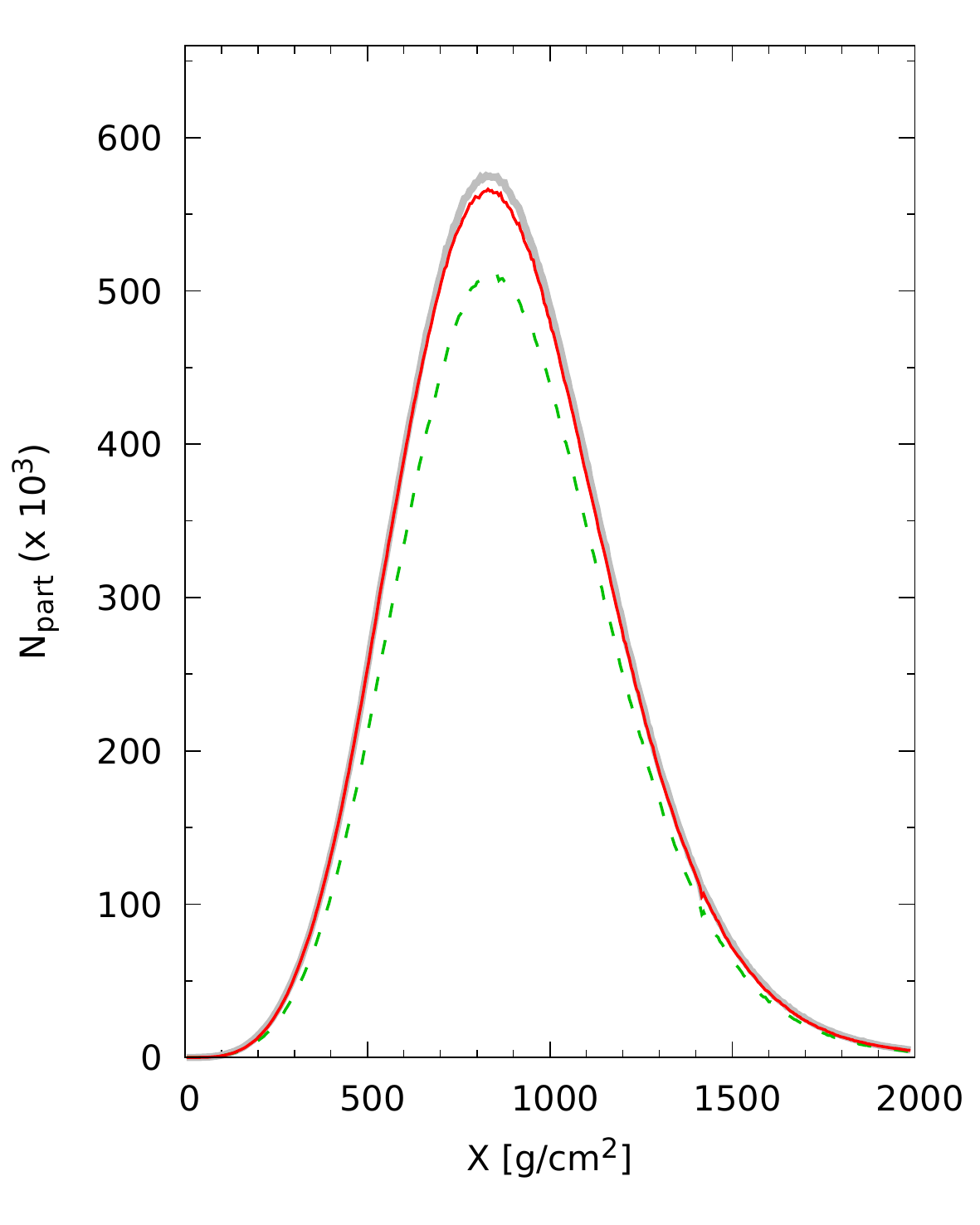} &
			\includegraphics[width=7cm]{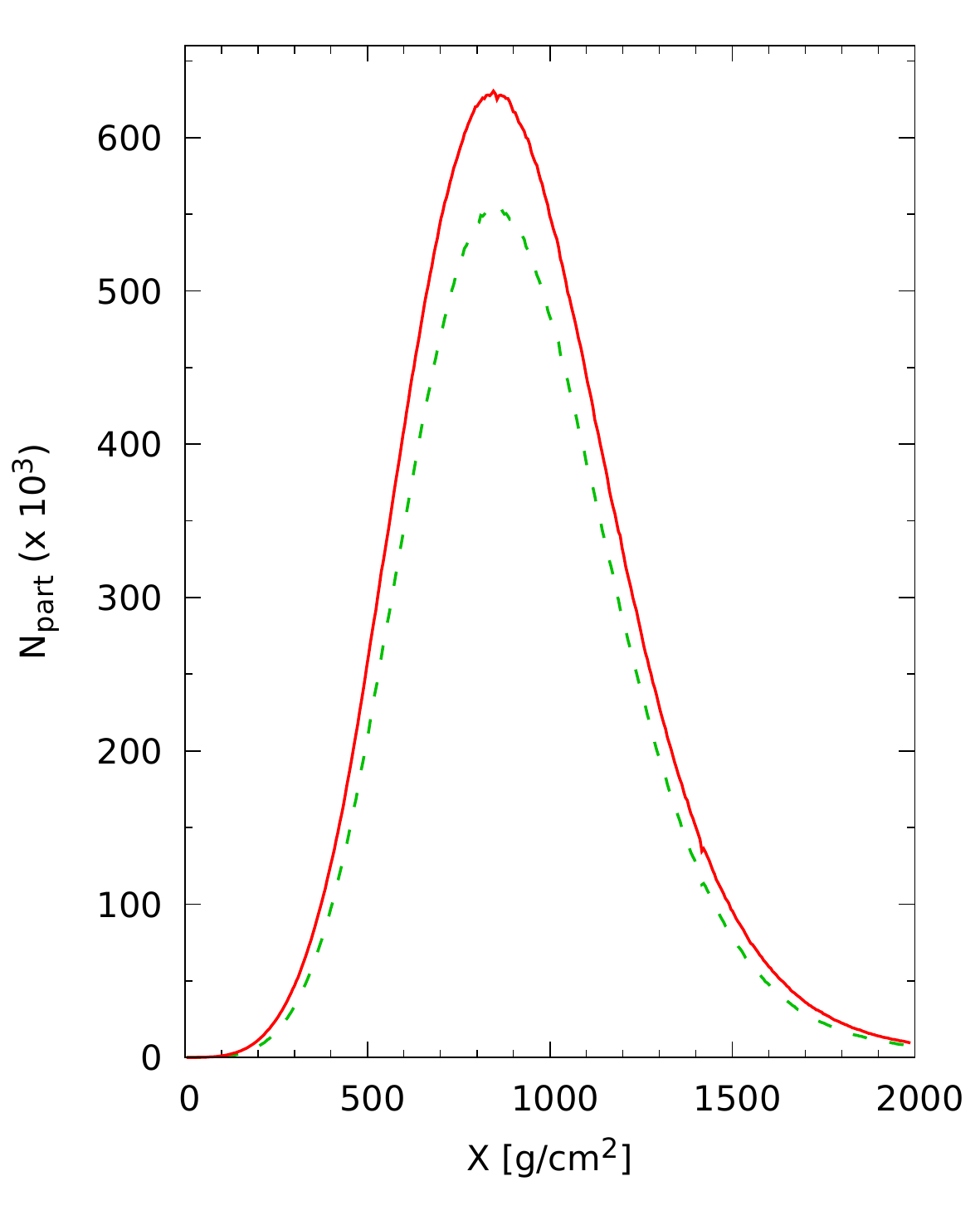}
			\\
			\includegraphics[width=7cm]{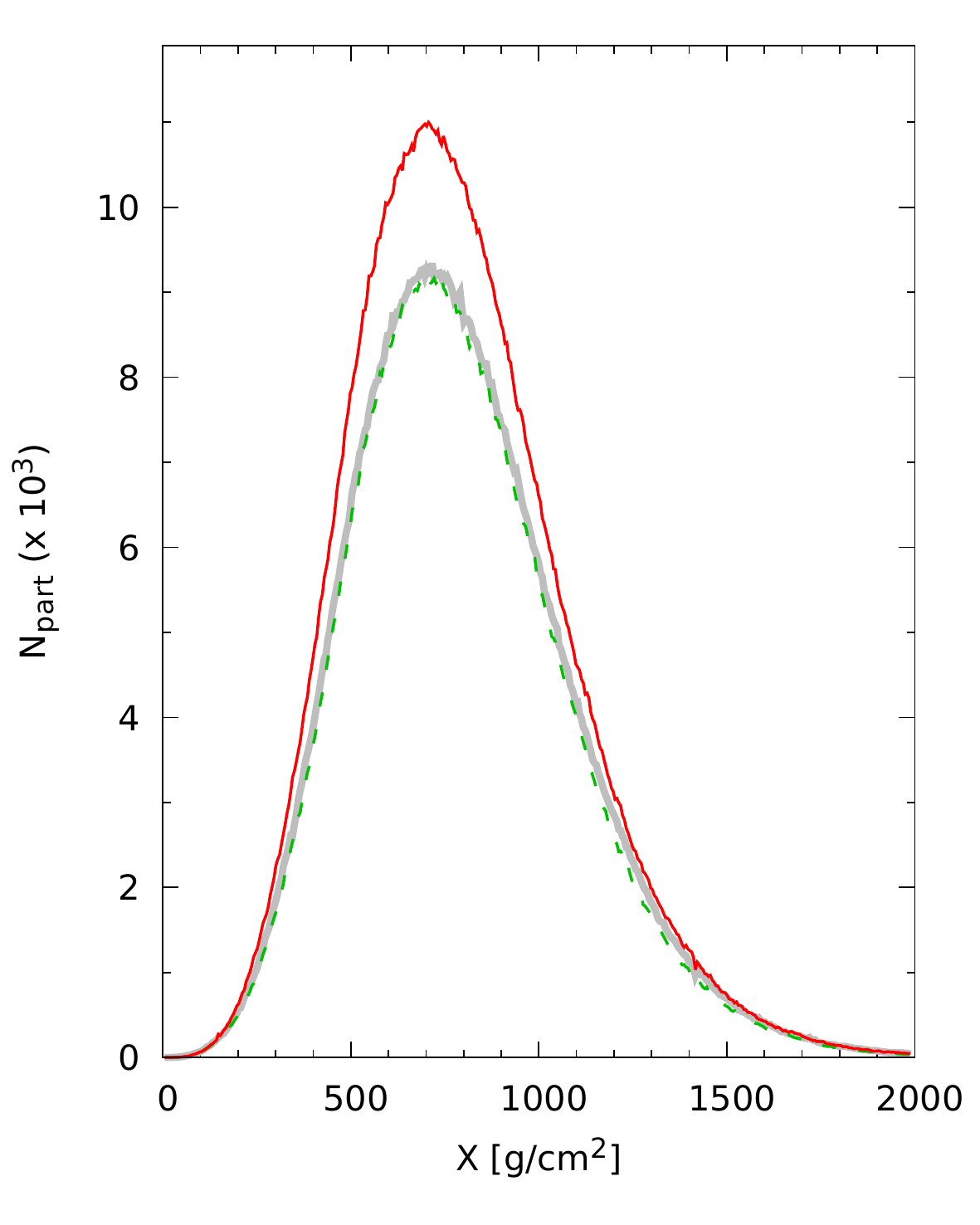} &
			\includegraphics[width=7cm]{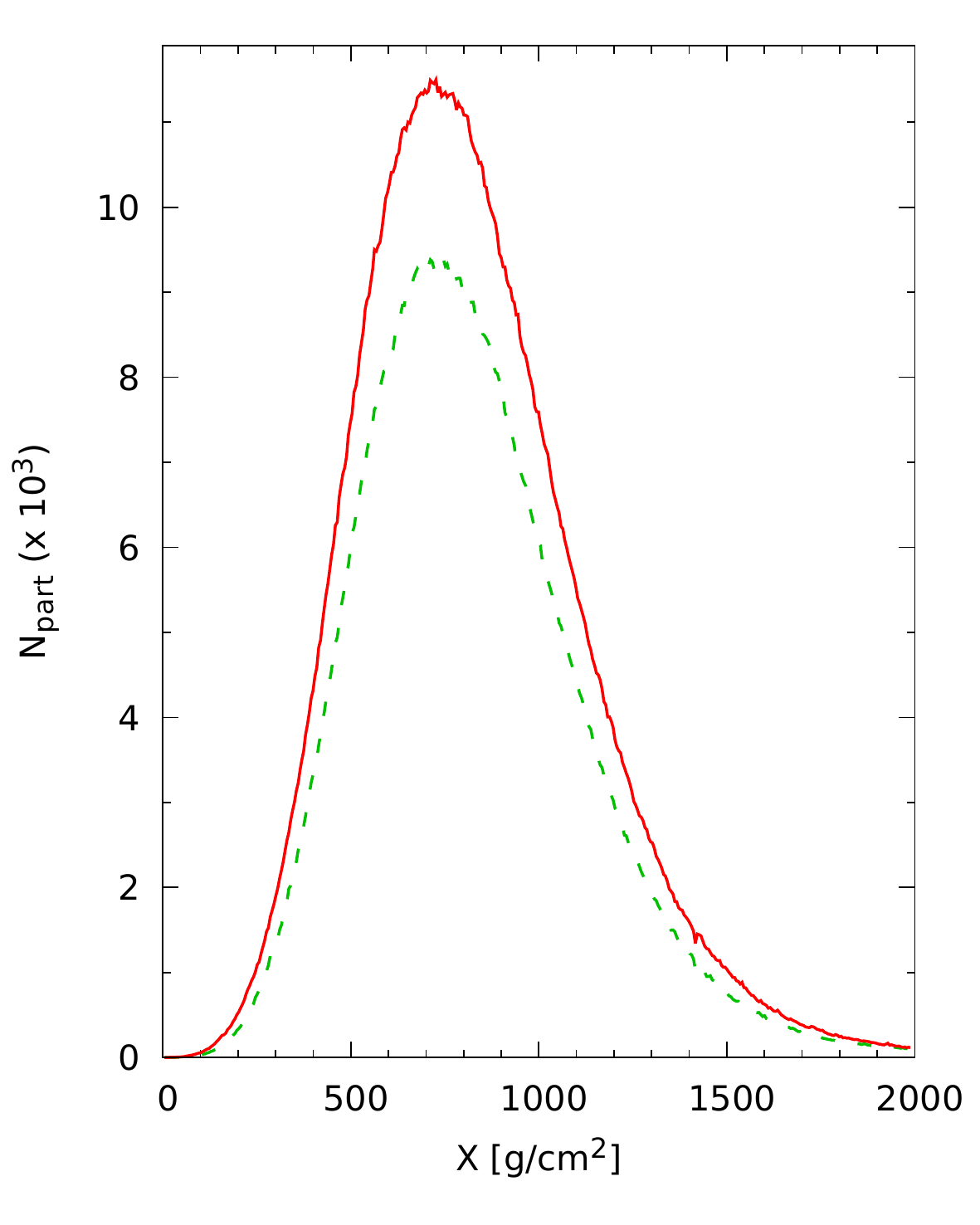}
		\end{tabular}
		\caption{Longitudinal development of pions (upper row) and kaons (lower   row) for $10^{19}$ eV photon
			showers inclined 80 degrees.
			The solid (dashed) lines correspond to simulations with the
			present BN-$\gamma$ (BH) model for photonuclear cross section, and processing high energy hadronic interactions with the QGSJET-II-04 (left) and EPOS (right) models. The grey line corresponds to similar simulations performed using the BN-$\gamma$ model for photonuclear cross section and the (pre-LHC) QGSJET-II-03 model (see figure 6 of reference \cite{Cornet:2015qda})}
		\label{fig:longipika}
	\end{center}
\end{figure}

\begin{figure}[tb]
  \begin{center}
	\begin{tabular}{cc}
	   \includegraphics[width=7cm]{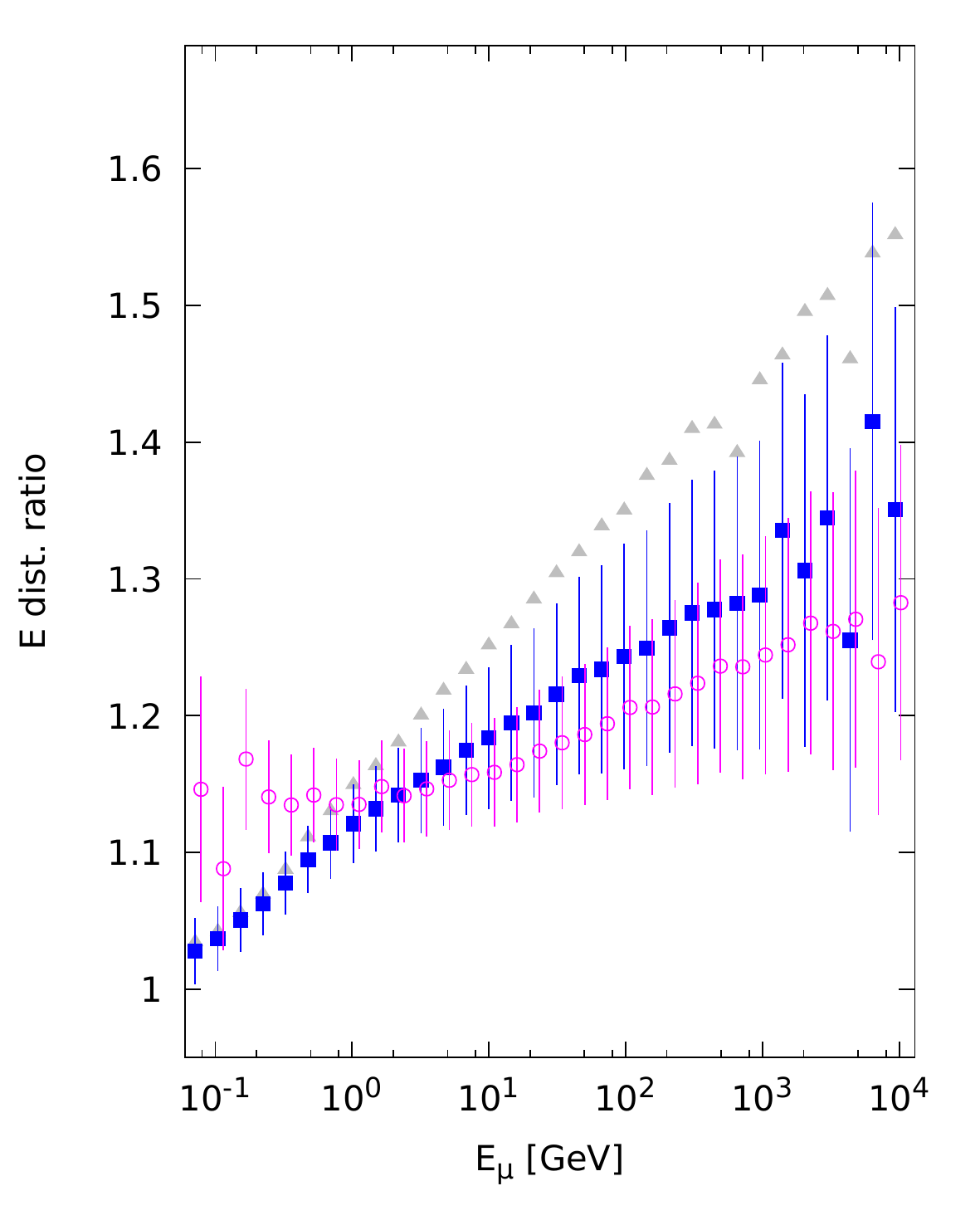} &
	   \includegraphics[width=7cm]{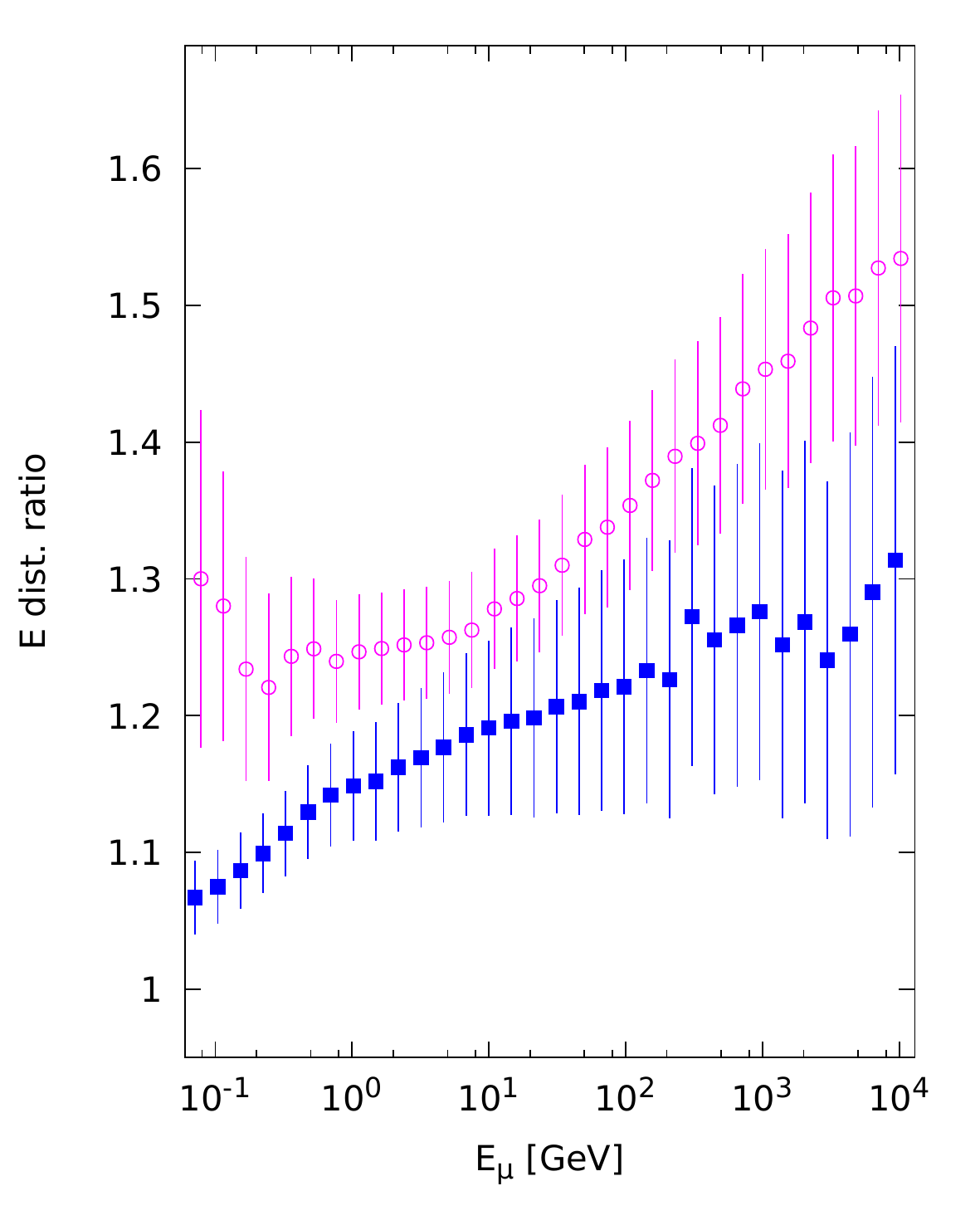}
	\end{tabular}
	\caption{Ratio between ground muon energy distributions obtained with
		the present (BN-$\gamma$) and old (BH) models, for $10^{19}$ eV photon showers, and simulating the hadronic simulations with QGSJET-II-04 (left) and EPOS (right). The solid squares (open circles) correspond to a shower inclination of 45
		(80) deg. Error bars are calculated by propagation of the individual RMS
		statistical errors of each of the distributions. The abscissas of the 80
		deg data set have been shifted by
		10\% to improve error bar visibility. The grey triangles correspond to similar simulations performed using the present BN-$\gamma$ model for photonuclear cross section and the (pre-LHC) QGSJET-II-03 model for showers inclined 45 degrees (see figure 10 of reference \cite{Cornet:2015qda}).}
    \label{fig:muegydistratio4580}
  \end{center}
\end{figure}

We have performed simulations of extended air showers using the AIRES
system \cite{AIRES} linked to the packages QGSJET-II-04 \cite{QGSJET-II-04} and EPOS 3.40 \cite{EPOS-LHC} for processing high energy hadronic interactions. The versions used for both hadronic models are those optimized taking into account the recent results of LHC experiments (for details see, for example, references \cite{EPOS-LHC,OstapchenkoLHC16}).

With each of the mentioned hadronic models we have run two sets
of simulations, namely, (1) using the cross sections for photonuclear
reactions at energies greater than 200 GeV that are provided with the
currently public version of AIRES which correspond to the lower of the two BH fits in the right panel of Figure~\ref{fig:sigjet_sigtot_gamp} (model BH introduced in Section~ \ref{sec:model});
%sect:totphotosigma});
 and (2) replacing those cross
sections by the ones corresponding to the present model \cite{Godbole:2008ex} (model BN-$\gamma$ introduced in Subsection~ \ref{ssec:BN}).

Following the lines of our previous work \cite{Cornet:2015qda}, we report here on the results for the
very representative case of $10^{19}$ eV gamma-initiated showers. As already pointed out in \cite{Cornet:2015qda}, at this
primary energy, geomagnetic conversion \cite{Billoir2001} is not
frequent, thus allowing photons to initiate normally the atmospheric shower development. 

In Figures~ \ref{fig:longimuons} and \ref{fig:compEPOSQGS} the results for the longitudinal development of the mean number of muons is displayed, while in Figure~\ref{fig:longipika} we present the results for the cases of pions and kaons.

For all the secondary particles considered, and for both hadronic models QGSJET and EPOS, the production of muons and hadrons is larger in the case of the simulations using the   BN-$\gamma$ model for photon cross sections, in comparison with the corresponding BH model  results, as expected (see discussion in reference \cite{Cornet:2015qda}). This shows up clearly in Figure~\ref{fig:longimuons} where the plots corresponding to the present model (solid red lines) are located always over those corresponding to the BH model (dashed green lines).

It is very important to notice that the number of hadrons produced during the shower development depends noticeably on the package used to process the hadronic collisions. Despite the fact that both hadronic packages have been tuned to best reproduce measurements performed at the LHC collider \cite{EPOS-LHC,OstapchenkoLHC16}, it is evident from the plots presented here that differences between models still persist. For example, the longitudinal profiles obtained using EPOS contain larger number of hadrons when compared with the corresponding ones for the case of QGSJET. Using model BN-$\gamma$  for photon cross sections (red solid lines in Figure~\ref{fig:longipika}), the number of pions at the point of maximum development is 11\% larger for EPOS with respect to QGSJET. In the case of kaons such figure is only 1\%, and for protons (neutrons) (not plotted) at the point of maximum development is 11\% (8\%) larger for EPOS-LHC with respect to QGSJET. As shower muons are generated after the decay of unstable hadrons, a similar increase can be seen for the maximum number of muons (Figure~\ref{fig:longimuons}) where EPOS overpasses the prediction of QGSJET-II in 6\%, as exemplified   in Figure~\ref{fig:compEPOSQGS}   for the case of muons when using model BN-$\gamma$ for photon cross section.

Another point that is important to check is the difference between pre and post-LHC versions of each model. In Figure~
%\ref{fig:longimuons}--
\ref{fig:longipika} we have included (grey lines in left column plots) the results corresponding to the present  model for photon cross sections, but simulated using QGSJET-II-03 (pre-LHC version of QGSJET). These grey lines correspond to the respective solid red lines plotted in figures 5 and 6 of reference \cite{Cornet:2015qda}. It can be observed that there are virtually no differences between the pre and post-LHC models in all the cases, except, remarkably, in the case of kaons (Figure~\ref{fig:longipika} lower left plot), where the post-LHC model QGSJET-II-04 predicts a noticeably larger number of kaons.

Another quantity that we have included in our study is the 
ratio of energy ditributions of muons reaching ground level (see the discussion on this quantity at reference \cite{Cornet:2015qda}). Such ratios are
plotted in Figure~\ref{fig:muegydistratio4580} as functions of the muon energy for the representative cases
of 45 (solid squares) and 80 (open circles) degrees of
inclination. The increased number of high energy muons resulting after
the simulations using the BN-$\gamma$ model for photonuclear cross
sections shows up clearly. In the case of QGSJET (left) the increase of the number of muons is not much dependent on the inclination of the shower, and is noticeably smaller compared with the results corresponding to QGSJET-II-03 (grey triangles), already studied in reference \cite{Cornet:2015qda}. In the case of EPOS-LHC (right) the results are somewhat different with respect to the QGSJET ones. In this case the very inclined showers (80 degrees) present the largest increase rate, reaching nearly 50\% for the most energetic muons, while for the showers inclined 45 degrees, the distribution ratios are about 25\% for the most energetic muons.

The differences between the results obtained in our analysis using different hadronic models, indicate that the simulation of hadronic collisions at the highest energies continue to be an open issue, more than 30 years after the first simulations were reported, and despite all the experimental data that have been collected since then.

\section*{Acknowledgements}

G.P. gratefully  acknowledges hospitality at the MIT Center for Theoretical Physics. This work was partially supported by CONICET and ANPCyT, Argentina.
 A.G. acknowledges partial support by the Ministerio de Econom\'\i a y Competitividad (Spain) under grant number FPA2016-78220-C3-3-P, 
and by Consejer\'\i a de Econom\'\i a, Innovaci\'on, Ciencia y Empleo, Junta de Andaluc\'\i a (Spain)(Grants FQM 101 and FQM 6552).
F. C. also acknowledges support by the Ministerio de Econom\'\i a y Competitividad (Spain) under grant number FPA 2016-78220-C3-1-P, 
and by Consejer\'\i a de Econom\'\i a, Innovaci\'on, Ciencia y Empleo, Junta de Andaluc\'\i a (Spain)(Grants FQM 330 and FQM 6552).

%We wish to thank {\bf to all : insert funding agencies}.


\begin{thebibliography}{99}
\bibitem{Cornet:2015qda}	
%Photoproduction total cross section and shower development
F. Cornet, C. A. Garcia Canal, A. Grau, G. Pancheri and S.J. Sciutto, {\it  Phys. Rev. } {\bf D92} (2015) 114011.
%DOI: 10.1103/PhysRevD.92.114011
%e-Print: arXiv:1510.07279
\bibitem{AIRES} S. J. Sciutto,
 Proceedings of the 27th International Cosmic Ray Conference, Hamburg, 2001,\\
 p. 237; %\hfil\break
  see also \url{http://www2.fisica.unlp.edu.ar/aires}.
\bibitem{QGSJET-II-04}
S. Ostapchenko, {\it Phys. Rev.} {\bf D83}, (2011) 014018;  %arxiv:1010.1869 S. Ostapchenko, 
{\it Phys. Rev.} {\bf D81}, (2010) 114028. %,arxiv:1003.0196
\bibitem{EPOS-LHC}
T. Pierog, I. Karpenko, J.M. Katzy, E. Yatsenko and K. Werner, {\it Phys. Rev.}
{\bf C92} (2015) 034906. %, arxiv:1306.0121
\bibitem{Block:2004ek} 	
%Evidence for the saturation of the Froissart bound
M.M. Block and F. Halzen %(Wisconsin U., Madison)May 2004. 6 pp.Published in 
{\it Phys. Rev.} {\bf D70} (2004) 091901.
%, NUHEP-989, MADPH-04-1382DOI: 10.1103/PhysRevD.70.091901
%e-Print: hep-ph/0405174
\bibitem{Godbole:2008ex}	
%Total photoproduction cross section at very high energy
R.M. Godbole, A. Grau, G. Pancheri and Y.N. Srivastava, {\it  Eur. Phys. J. C} {\bf 63} (2009) 69-85.
%e-Print: arXiv:0812.1065
\bibitem{Godbole:2004kx}
%Soft gluon radiation and energy dependence of total hadronic cross sections
R.M. Godbole, A. Grau, G. Pancheri and Y.N. Srivastava, {\it Phys. Rev. D} {\bf 72}  (2005) 076001.
%e-Print: hep-ph/0408355
\bibitem{Grau:1999em}
%Hadronic total cross sections through soft gluon summation in impact parameter space
A. Grau, G. Pancheri and Y.N. Srivastava, %May 1999. 25 pp.
{\it Phys. Rev.} {\bf D60} (1999) 114020.
%UG-FT-96-99, LNF-99-010-P, LNF-99-010(P)
%DOI: 10.1103/PhysRevD.60.114020
%e-Print: hep-ph/9905228
\bibitem{Durand:1988ax} 	
%Semihard (QCD) and High-energy pppp and p???ppp Scattering
Loyal Durand and Hong Pi, {\it Phys. Rev.} {\bf D40} (1989) 1436.
\bibitem{GRS} M. Gl\"uck, E. Reya and I. Schienbein,
  {\it Phys. Rev.} {\bf D60} (1999) 054019;
  Erratum:{\it Phys. Rev.} {\bf D62} (2000) 019902.
 \bibitem{GRV} M.~Gl\"uck, E.~Reya, and A.~Vogt,
  {\it Z. Phys.} {\bf C53} (1992) 127--134;
  {\it Z. Phys.} {\bf C67} (1995) 443--448;
  {\it  Eur. Phys. J.} {\bf C5} (1998) 461--470.
 \bibitem{MRST} A.~D. Martin, R.~G. Roberts, W.~J. Stirling, and
  R.~S. Thorne, {\it Phys. Lett.} {\bf B531} (2002) 216--224.
\bibitem{Donnachie:1992ny} A. Donnachie and P. V. Landshoff, {\it Phys.Lett.}  {\bf B296} (1992) 227-232.
\bibitem{Bloch:1937pw}
%	Note on the Radiation Field of the electron
F. Bloch, A. Nordsieck %(Stanford U., Phys. Dept.). Jul 1937. 6 pp.
{\it Phys. Rev.} {\bf 52} (1937) 54-59.
\bibitem{Corsetti:1996wg} 
%Bloch-Nordsieck summation and partonic distributions in impact parameter space
A. Corsetti, A. Grau, G. Pancheri and Y.N. Srivastava, {\it Phys. Lett.} {\bf B382} (1996) 282-288.
\bibitem{Fagundes:2017xli}	
%The Inelastic cross section and Survival Probabilities at LHC in mini-jet models
D. A. Fagundes, A. Grau, G. Pancheri, O. Shekhovtsova and Y. N. Srivastava,
%e-Print: arXiv:1706.00093. 
{\it Phys. Rev.} {\bf D96} (2017) 054010. 
\bibitem{Greco} P. Chiappetta and M. Greco, {\it Nucl. Phys.} {\bf B199} (1982) 77.
\bibitem{Fletcher:1992mw} %QCD and minimum bias physics: The Importance of HERA photoproduction measurements
R.S. Fletcher, T.K. Gaisser and F. Halzen, {\it Phys. Lett.} {\bf B298} (1993) 442-444.
%\bibitem{Donnachie:1992ny} A. Donnachie and P. V. Landshoff, {\it Phys.Lett.}  {\bf B296} (1992) 227-232.
%\bibitem{EPOS-LHCtest}
%EPOS LHC : test of collective hadronization with LHC data
%\textcolor{red}{T. Pierog, Iu. Karpenko, J.M. Katzy, E. Yatsenko, K. Werner, Phys. Rev. C 92, 034906 (2015).}
\bibitem{OstapchenkoLHC16}
%LHC results and hadronic interaction models.
S. Ostapchenko,
Proc. of the XXV European Cosmic Ray Symposium, Turin (2016),
eConf C16-09-04.3, arXiv:1612.09461v1 [astro-ph.HE].
\bibitem{Billoir2001} P. Billoir {\it et al.,\/} for The Pierre Auger Collaboration, 
Proceedings of the 27th International Cosmic Ray Conference, Hamburg, 2001, p. 718.
%{\it Proc. 27th ICRC (Hamburg)}{1}{718}{2001}.
     \end{thebibliography}
\end{document}